\documentclass[twocolumn]{aa}
\usepackage{epsfig}
\def\simlt{\ \raise -2.truept\hbox{\rlap{\hbox{$\sim$}}\raise5.truept   %
\hbox{$<$}\ }}
\def\simgt{\ \raise -2.truept\hbox{\rlap{\hbox{$\sim$}}\raise5.truept   %
\hbox{$>$}\ }}                                                          %

\def\be{\begin{equation}}
\def\ee{\end{equation}}
\def\newline{\hfil\break}

\def\la{\mathrel{\hbox{\rlap{\hbox{\lower4pt\hbox{$\sim$}}}\hbox{$<$}}}}
\def\ga{\mathrel{\hbox{\rlap{\hbox{\lower4pt\hbox{$\sim$}}}\hbox{$>$}}}}

\begin{document}

\title{The impact of magnetic field on the cluster M-T relation}

   \author{S. Colafrancesco  \& F. Giordano  }

   \offprints{S. Colafrancesco}

\institute{   INAF - Osservatorio Astronomico di Roma,
              via Frascati 33, I-00040 Monteporzio, Italy.\\
              Email: cola@mporzio.astro.it
             }

\date{Received 11 April 2006 / Accepted 09 June 2006 }

\authorrunning {S. Colafrancesco \& F. Giordano}

\titlerunning {Magnetic field and the cluster M-T relation}

\abstract{We discuss the impact of magnetic field on the mass -- temperature relation for
groups and clusters of galaxies based on the derivation of the general Magnetic Virial
Theorem. The presence of a magnetic field $B$ yields a decrease of the virial temperature
$T$ for a fixed mass $M$: such a decrease in $T$ is stronger for low-mass systems than
for high-mass systems. We outline several implications of the presence of $B$-field and
of its mass scaling for the structure and evolution of groups and clusters.

}

 \maketitle


\section{Introduction}
 \label{intro}

Magnetic fields fill intracluster and interstellar space, affect the evolution of
galaxies, contribute significantly to the total pressure of interstellar gas, are
essential for the onset of star formation, and control the diffusion, the confinement and
the evolution of cosmic rays in the interstellar and intracluster medium (ICM).
In clusters of galaxies, magnetic fields may play also a critical role in regulating heat
conduction (e.g., Chandran et al. 1998, Narayan \& Medvedev 2001), and may also govern
and trace cluster formation and evolution.\\
We know that magnetic fields exist in clusters of galaxies for several reasons.
First, in many galaxy clusters we observe the synchrotron radio-halo emission produced by
relativistic electrons spiraling along magnetic field lines.
Second, the Faraday rotation of linearly polarized radio emission traversing the ICM
proves directly and independently the existence of intracluster magnetic fields (see,
e.g., Carilli \& Taylor 2002, Govoni \& Feretti 2004 for recent reviews).
The Rotation Measure (RM) data throughout the inner ($\sim 0.5$ Mpc) cluster region
support magnetic field strengths of the order of several to tens of $\mu$G (see Carilli
\& Taylor 2002, Govoni \& Feretti 2004).
The high local values of B observed in the central, cool region of clusters are likely
related, however, to quite special conditions (such as turbulent amplification of the
local $B$-field driven by radio bubbles or AGN jets, see e.g. Ensslin \& Vogt 2006) and
thus are probably not representative of the overall system (see, e.g. Carilli \& Taylor
2002).
Other estimates of the magnetic field strength on the cluster wide scale come from the
combination of synchrotron radio and inverse Compton detections in the hard X-rays (e.g.,
Colafrancesco, Marchegiani \& Perola 2005), from the study of cold fronts and from
numerical simulations (see, e.g., Govoni \& Feretti 2004).
This evidence provides indication on the wide-scale B-field which is at the level of a
few tens up to several $\mu$G (and in some cases up to $\sim 10$ $\mu$G, as in Coma) with
the larger values being attained by the most massive systems.\\
Numerical simulations (e.g., Dolag et al. 2001a) have shown that the wide-scale magnetic
fields in massive clusters produce variations of the cluster mass at the level of $\sim
5-10 \%$ of their unmagnetized value. Such mass variations induce a comparable variation
on the IC gas temperature $T$ for virialized systems. Such variations are not expected to
produce strong variations in the relative $M-T$ relation for massive clusters.\\
The $M-T$ relation predicted in a pure CDM model for $B=0$ follows the self-similar
scaling $M \propto T^{\eta}$ with $\eta = 3/2$ (see, e.g., Colafrancesco et al. 1997,
Arnaud 2005).
A Chandra study (Allen et al. 2001) of five hot clusters (with $k_B T_g > 5.5$ keV)
derived a $M-T$ relation slope of $\eta = 1.51 \pm 0.27$, consistent with the
self-similar model. However, due to the relatively small Chandra field of view, the $M-T$
relation was established at $R_{2500}$, i.e., about $0.3 R_{200}$ (here $R_{\delta}$ and
$M_{\delta}$ are the radius and mass at which the density contrast of the system is
$\delta$). More recently, the $M-T$ relation was established down to lower density
contrasts ($\delta = 200$) from a sample of ten nearby relaxed galaxy clusters covering a
wider temperature range, $k_B T_g \approx 2 - 9$ keV (Arnaud et al. 2005). The masses
were derived from mass profiles measured with XMM-Newton at least down to $R_{1000}$ and
extrapolated beyond that radius using the NFW (Navarro, Frenk \& White 1997) model. The
$M_{2500}-T$ for hot clusters is consistent with the Chandra results. The slope of the
$M-T$ relation is the same at all $\delta$ values, reflecting the self-similarity of the
mass profiles. At $\delta= 500$ the slope of the relation for the sub-sample of hot
clusters ($k_B T_g> 3.5$ keV) is $\eta = 1.49 \pm 0.15$ consistent with the standard CDM
self-similar expectation. The relation, however, steepens when the whole sample of
clusters is considered, providing a slope $\eta = 1.71 \pm 0.09$. The normalisation of
the $M-T$ relation differs, at all density contrasts from the prediction of pure
gravitation based models by $\sim 30\%$ (see Arnaud 2005 for a discussion).\\
In this Letter we will explore the effect of wide-scale magnetic fields on the $M-T$
relation over a large range of masses and temperatures by using the predictions of the
magnetic virial theorem. We will discuss its implications for the evolution and the
scaling relations of magnetized clusters.
The relevant physical quantities are calculated using $H_0 = 71$ km s$^{-1}$ Mpc$^{-1}$
and a flat, vacuum-dominated CDM  ($\Omega_{\rm m} = 0.3, \Omega_{\Lambda}=0.7$)
cosmological model.

\section{The magnetic virial theorem for galaxy clusters}
 \label{sect.mvt}

Under the assumption of a ICM in hydrostatic equilibrium with the potential well of a
spherically-symmetric, isolated, virialized and magnetized cluster, the general relation
between the ICM temperature $T$ and the cluster virial mass $M$ is obtained by applying
the magnetic virial theorem (MVT):
 \be
\frac{1}{2}\frac{d^2I_{ik}}{dt^2}=2K_{ik}+\frac{2}{3}U \delta_{ik}+\int_VF_{ik}d^3x
+W_{ik}\label{th.viriale} \, ,
 \ee
where $I_{ik}$ is the inertia momentum tensor, $K_{ik}$ is the kinetic energy tensor, $U$
is the thermal energy of the intra-cluster gas, $F_{ik}$ is the Maxwell tensor associated
to the magnetic field and $W_{ik}$ is the potential energy tensor. The full derivation of
the MVT is reported in the Appendix. For a static and isothermal galaxy cluster the trace
of eq. (\ref{th.viriale}) yields the condition
 \be
 2K+2U + U_B + W = 0 \, ,
 \label{eq.mvt}
 \ee
where $U_B$ is the magnetic energy of the system, $U$ is the kinetic energy of the gas,
$K$ is the dark-matter particle kinetic energy and $W$ is the potential energy of the
system (see Appendix for details). The previous eqs. (\ref{th.viriale}) and
(\ref{eq.mvt}) hold specifically in the absence of an external medium.
For the general case of a cluster which is immersed in a Inter Galactic Medium (IGM) or
external medium which exerts an external pressure $P_{ext}$, eq.(\ref{eq.mvt}) yields the
formula for the temperature of the gas in virial equilibrium
\be
 \frac{k_BT_g}{\mu m_p}=\frac{ \xi G}{3}\frac{M_{vir}}{r_{vir}}
 \left(1-\frac{M_{\phi}^2}{M_{vir}^2}+
 \frac{4\pi}{\xi G}\frac{r_{vir}^4}{M_{vir}^2}P_{ext}\right),
 \label{eq.TvirM}
 \ee
where usually $\xi \simgt 1$ and we defined the quantity
 \be
 M_{\phi}\simeq 1.32\cdot 10^{13} M_{\odot} \left[\frac{I(c)}{c^3} \right]^{1/2}
 \left(\frac{B_*}{\mu G}\right)\left(\frac{r_{vir}}{Mpc}\right)^2 \, ,
 \label{eq.mb}
 \ee
where $I(c)=\int_0^c (\rho_g(r=0)/{\bar \rho_g}(z=0))^{2\alpha} x^2 y_g^{2\alpha}(x,B=0)
dx$. Here $c= r_{vir}/r_s$ (we assume a NFW Dark Matter density profile with scale radius
$r_s$) and $y_g(x,B=0) = \rho_g(x)/\rho_g(x=0)$ is the gas density profile normalized to
the central gas density (i.e. the solution of the hydrostatic equilibrium equation in the
absence of magnetic field, see Colafrancesco \& Giordano 2006a for details). The radial
profile of the magnetic field has been assumed as $B(r) = B_* [\rho_g(r,B)/10^4 {\bar
\rho}(z=0)]^{\alpha}$ with $\alpha=0.9$ (see, e.g., Dolag et al. 2001b).\\
For the case $P_{ext}=0$ and $B=0$, the quantity $M_{\phi}=0$ and the well-known relation
 \be
 k_B T_g(B=0)=-{ \xi \mu m_p W \over 3 M_{vir} }
 \label{eq.t_b0}
 \ee
re-obtains (here $\mu=0.63$ is the mean molecular weight, corresponding to a hydrogen
mass fraction of $0.69$, $m_p$ is the proton mass and $k_B$ is the Boltzmann constant).\\
For $B > 0$, the quantity $M_{\phi} > 0$ and the gas temperature at fixed $M_{vir}$, as
obtained from eq.(\ref{eq.TvirM}), is
 \be
 k T_g = k T_g(B=0) \left(1-\frac{M_{\phi}^2}{M_{vir}^2}+
 \frac{4\pi}{\xi G}\frac{r_{vir}^4}{M_{vir}^2}P_{ext}\right) \, ,
 \label{eq.TvirM_norm}
 \ee
and is lower (for $P_{ext}=0$) than that given by eq.(\ref{eq.t_b0}) because the
additional magnetic field energy term $U_B$ adds to the MVT. The presence of an external
pressure $P_{ext}$ tends to compensate the decrease of $T_g$ induced by the magnetic
field.
For values of the temperature and density of the IGM (as estimated by the WHIM structure
around large-scale overdensities, see, e.g., Fang \& Bryan 2001), $P_{ext} \sim
1.7\cdot10^{-3}$ eV cm$^{-3} (n_{IGM}/10^{-5} cm^{-3})(T_{IGM}/2 \cdot 10^6 K)$.
However, in the outer regions of massive clusters (at $r \simgt r_{vir}$) the external
gas pressure can reach values $P_{ext} \sim 0.2$ eV cm$^{-3} (n/10^{-4} cm^{-3})(T_g/1.7
\cdot 10^7 K)$ [here we considered the mean projected temperature profile for the cluster
sample studied by Piffaretti et al. (2005, see their Fig.4) and a typical cluster with
$T_X = 10$ keV]. In such a case, the value of $P_{ext}$ is a significant fraction $\sim 4
\%$ of the central ICM pressure and $\sim 50 \%$ of the ICM pressure at the virial radius
for a typical cluster. Thus, it cannot be neglected in the $T_g$ estimate from
eq.(\ref{eq.TvirM_norm}).
A value $P_{ext} \sim 0.2$ eV cm$^{-3}$, as estimated at the outskirts ($r \simgt
r_{vir}$) of rich clusters, can be considered as an upper bound to $P_{ext}$, since an
exact determination of the total cluster mass (which is subject to various systematic
uncertainties, see, e.g., Rasia et al. 2006) certainly requires to go beyond $r_{vir}$.
We thus consider in the following this value of $P_{ext}$ as a reference upper bound to
be used in our temperature estimate from eq.(\ref{eq.TvirM_norm}) in the presence of a
B-field. Lower values of $P_{ext}$, down to its value in the WHIM, have progressively
minor importance.\\
For reasonable values of $B_* \simgt$ a few $\mu$G, the quantity $M^2_{\phi}
> M_{vir}^2 \cdot (P_{ext}/P_{vir})$
(with $P_{vir} = (\frac{4\pi}{\xi G}\frac{r_{vir}^4}{M_{vir}^2})^{-1}$) in eq.(\ref{eq.TvirM}),
and the main effect is a reduction of the cluster temperature which is more pronounced
for less massive systems, where $ M_{\phi}$ becomes comparable to $M_{vir}$. The effect
of the magnetic field and of the external pressure are larger for low-$M$ clusters (see
Fig.\ref{fig.MT_B_P0}).

\section{The magnetized M-T relation}
 \label{sect.2}

\begin{figure}[h!]
\begin{center}
 \epsfig{file=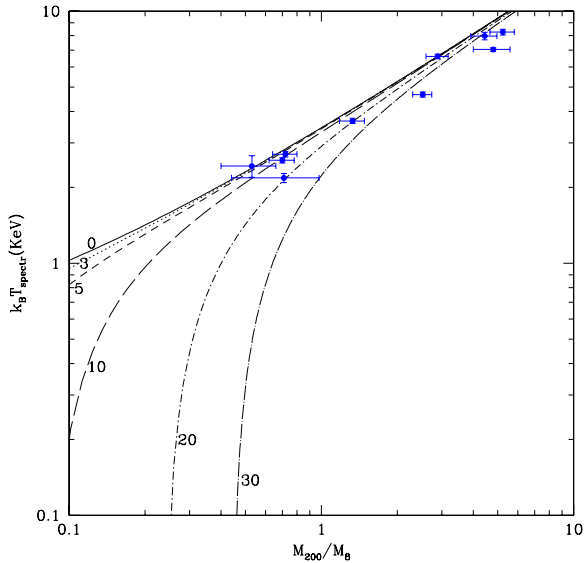,height=8.cm,width=8.cm,angle=0.0}
  \caption{\footnotesize{We show the $T_{spectr}-M_{200}$ relation at $z=0$ for clusters
  which contain a magnetic field $B_*$ in the illustrative range $0-30$ $\mu$G (as labelled)
  in the case of $P_{ext}=0.2$ eV cm$^{-3}$.
  Here $M_8 \simeq 2\cdot 10^{14} M_{\odot}h_{71}^{-1}$. Data are taken from Arnaud (2005).
  }}
  \label{fig.MT_B}
\end{center}
\end{figure}

The $T-M$ relation for magnetized clusters is shown in Fig.\ref{fig.MT_B}.
\begin{figure}[h!]
\begin{center}
 \epsfig{file=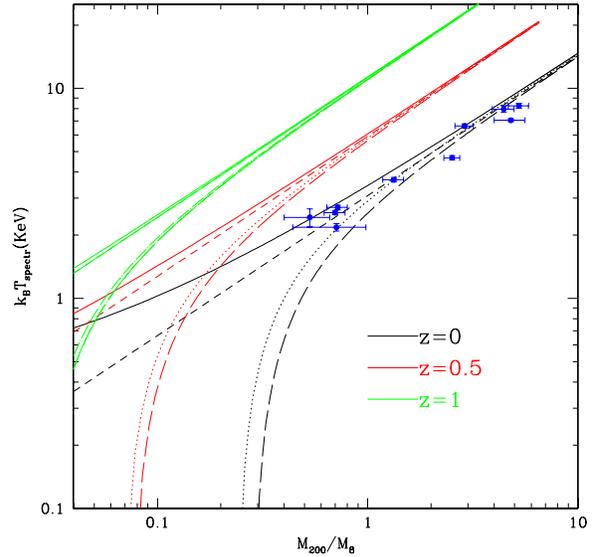,height=8.cm,width=8.cm,angle=0.0}
  \caption{\footnotesize{Same as Fig.\ref{fig.MT_B} but for a $P_{ext}=0$ with $B_*=0$
  (solid curves) and $B_*=20$ $\mu$G (long-dashes curves) and $P_{ext}=0.2$ eV cm$^{-3}$
  for the same values of $B_*=0$ (short-dashes curves) and $B_*=20$ $\mu$G (dotted curves)
  The evolution of the $T_{spectr}-M_{200}$ relation for magnetized clusters is shown at
  $z=0$ (black), $z=0.5$ (red) and $z=1$(green).
  }}
  \label{fig.MT_B_P0}
\end{center}
\end{figure}
We normalize the $T_{spectr.}-M_{200}$ relation for the case $B=0$ to the observed data
derived by Arnaud et al. (2005) by assuming $T_{spectr} = T_g$ with $\xi \approx 1.5$ in
eq.(\ref{eq.TvirM}), as can be expected from the continuous shock-heating of the IC gas
within the virial radius after the formation of the original structure (see, e.g., Makino
et al. 1998, Fujita et al. 2003, Ryu et al. 2003).
Such a prescription for $\xi$ has been used by the previous authors for unmagnetized
clusters in the absence of external pressure.
A systematic effect which goes towards the direction of increasing the cluster
temperatures is the presence of a minimal external pressure in eq.(\ref{eq.TvirM}). A
value $P_{ext} \sim(0.1-0.2) P_{vir}$ (like that found in the IGM around clusters) could
easily accommodate for an overall value of $ \xi \approx 1.5 \times  (P_{ext} / P_{vir})
> 1.5$ and reasonably in the range $1.65 - 1.8$, for the previous values of $P_{ext}$.
Given the large theoretical uncertainty on the non-gravitational heating efficiency, we
adopt an overall value $\xi \approx 1.8$ to normalize our prediction to the data point at
$M_{200}  \approx 3 \cdot M_8$ in Fig.1, which is the point with the smaller intrinsic
error. Values of $\xi$ in the plausible range $1.5 - 1.8$ marginally change, however, our
predictions.\\
The relation $M_{200} \simeq 0.77 M_{vir}$ is also found in our mass scale definition.\\
Small variations of temperatures with respect to their unmagnetized values are found for
massive clusters since the quantity $M_{\phi} \ll M_{vir}$ in this mass range and the
value of $P_{ext}$ has little or negligible effect (see Fig.\ref{fig.MT_B_P0}).
This is in agreement with the results of numerical simulations (Dolag et al. 2001a).
However, when $M_{\phi}$ becomes comparable to $M_{vir}$, the IC gas temperature becomes
lower than its unmagnetized value and the $T-M$ relation steepens in the range of less
massive systems like groups and poor clusters. The temperature $T_g$ formally tends to
zero when $M_{\phi} \to M_{vir}(1+P_{ext}/P_{vir})^{1/2}$. However, this limit is
unphysical since it corresponds to an unstable system in which the magnetic pressure
overcomes the gravitational pull. Thus, any physical configuration of magnetized
virialized structures must have $M_{\phi} < M_{vir}(1+P_{ext}/P_{vir})^{1/2}$.
The effect of $P_{ext}$ counterbalances the effect of the $B$-field on the $T-M$
relation, increases for low-$M$ systems and decreases with increasing redshift (see
Fig.\ref{fig.MT_B_P0}) because $P_{vir}$ increases with increasing redshift.

\section{Discussion and conclusions}
 \label{conclusion}

We have derived here, for the first time, a relation between the temperature of the IC
gas from the general MVT in the presence of magnetic field and external pressure. The
result of the MVT for clusters bring relevant modifications to the gas temperature for
virialized and magnetized clusters.
As a consequence, the observed $T-M$ relation is steeper than the simple predictions of a
$\Lambda$CDM structure formation scenario and its effective slope increases in the
low-$M$ region.
However, since the masses of the observed clusters have been derived under the assumption
of absence of $B$ field, the slope indicated by the data of the observed $T-M$ relation
could be not completely representative.
In this context we also stress that the $T-M$ relation might be affected by other
systematic uncertainties in the mass derived by X-ray observations (e.g., Rasia et al.
2006) which would change the slope of the $T-M$ relation especially in the low-$M$ range.
A robust analysis of the cluster mass estimate should require the use of a detailed
hydrostatic equilibrium condition in combination with reliable temperature profiles. In
both these aspects the effect of the B-field is relevant and should be taken into
account.
Furthermore, the predictions at low-$T$, where the effects of the $B$-field are stronger,
are rendered uncertain by the absence of a clear definition of a spectroscopic
temperature (e.g., Mazzotta et al. 2004). Thus, a complete analysis of the $T-M$ relation
relies on a very detailed understanding of the physical properties of the IC gas in the
presence of B-field with the input of a precise total mass reconstruction and temperature
determination.

The results we derived here have a broad range of implications on cluster structure and
evolution:
flattening of the entropy -- temperature relation and higher entropies in cluster cores
are expected in the presence of magnetic fields. Further effects on the X-ray luminosity
-- temperature relation are also expected as well as modifications of the thermal
Sunyaev-Zel'dovich effect.
Since these studies are far beyond the scope of this paper, we refer the interested
reader to much more detailed analysis which are presented elsewhere (Colafrancesco \&
Giordano 2006a,b).

To conclude, we notice that a full description of the structure and evolution of the
population of groups and clusters of galaxies which considers also the role of magnetic
fields will definitely shed light on several, still unclear aspects of the interference
between gravitational and non-gravitational mechanisms in the evolution of these systems,
and calls for a more refined physical description to use galaxy clusters as appropriate
cosmological probes.

\begin{acknowledgements}
The authors thank the Referee for useful comments. This work is supported by PRIN-MIUR
under contract No.2004027755$\_$003.
\end{acknowledgements}

\appendix
\section{The MVT for galaxy clusters}
 \label{App}

Let us introduce the following quantities:
\begin{eqnarray}
\varphi(r)&=&-G\int_V
\frac{\rho(x_j^{\prime})}{|x_j-x_{j'}|}d^3x^{\prime}\\
W&=&-\frac{1}{2}G\int_V\int_V\frac{\rho(x_j)\rho(x_j^{\prime})}{|x_j-x_{j'}|}
d^3x d^3x^{\prime}\\
F_{ij}&=&\frac{B^2}{8\pi}\delta_{ij}-\frac{B_iB_j}{4\pi}
\end{eqnarray}
where $\varphi$ is the gravitational potential, $W$ is the gravitational energy, $F_{ij}$
is the Maxwell tensor associated to the magnetic field and $\rho=\rho_{dm}+\rho_g$ is the
total density of the cluster (we neglect here the subdominant contribution of galaxies).
Then, the equation of motion for the systems given by the Euler equation writes as
\be
\rho\left(\frac{\partial}{\partial
t}+(\vec{v}\cdot\vec{\nabla})\right)v_i=-\frac{\partial
p}{\partial x_i}-\frac{\partial F_{ij}}{\partial
x_i}-\rho\frac{\partial\varphi}{\partial x_i}\label{eq.Eulero} \ee
where $v_i$ is the i-th component of the velocity and $p$ is the total pressure given by
$p_{dm}+p_g$. Here, $p\sim p_g$ since we assume that DM is cold and collisionless, i.e.
$p_{dm}\sim0$ (we consider a fluid with no viscosity for which $P_{ij}=p$ $\delta_{ij}$).
Multiplying by $x_k$ and integrating over the cluster volume we obtain:
\begin{eqnarray}
 &&\int_V x_k\frac{\partial}{\partial t}(\rho v_i)d^3x\nonumber\\
 &&+\int_Vx_k\frac{\partial}{\partial x_j}(\rho v_iv_j)d^3x= -\int_Vx_k\frac{\partial
p}{\partial x_i}d^3x\\
 &&-\int_Vx_k\rho\frac{\partial\varphi} {\partial x_i}d^3x-\int_V
x_k\frac{\partial F_{ij}}{\partial x_j}d^3x.\nonumber
\end{eqnarray}
Using the continuity equation and the standard integral theorems, we convert the first
member of this equation in the form
\begin{eqnarray}
\int_V x_k\frac{\partial}{\partial t}(\rho v_i)d^3x & + &
\int_Vx_k\frac{\partial}{\partial x_j}(\rho v_iv_j)d^3x\\
 &=& \frac{d}{dt}\int_V\rho
v_iv_jd^3x-2K_{ik}+\oint x_k \rho v_iv_jdS_j  \nonumber
\end{eqnarray}
where $K_{ij}=1/2\int\rho_{dm} v_iv_jd^3x$ indicates the dark-matter kinetic energy
tensor. The second member of eq.(A.5) writes as
\begin{eqnarray}
-\int_Vx_k\frac{\partial p}{\partial
x_i}d^3x&=&\frac{2}{3}U\delta_{ik}-\oint dS_jx_kp\label{Pext}\\
-\int_Vx_k\frac{\partial F_{ij}}{\partial
x_j}d^3x&=&\int_VF_{ik}d^3x-\oint dS_jF_{ij}x_k
\end{eqnarray}
where $U=3/2\int p_g~d^3x$ is the IC gas thermal energy and one can show that
 \be
 -\int_Vx_k\rho\frac{\partial \phi}{\partial x_i}d^3x=\frac{1}{2}W_{ik}.
 \ee
Neglecting (in our case) the surface integrals (these physical quantities are negligible
when the integration surface is chosen far from the cluster center) one obtains:
 \be
\frac{1}{2}\frac{d^2I_{ik}}{dt^2}=2K_{ik}+\frac{2}{3}U \delta_{ik}+\int_VF_{ik}d^3x
+W_{ik} \; ,
 \label{th.viriale1}
 \ee
where $I_{ij}$ is defined as
 \be
I_{ij}=\int_V\rho x_ix_jd^3x.
 \ee
Using the trace of eq.(\ref{th.viriale1}), we obtain the equation:
 \be
\frac{1}{2}\frac{d^2I}{dt^2}=2K+2U+U_B +W
 \label{th.viriale2}
 \ee
where $U_B\equiv F$, and $F\equiv\int\frac{B^2(r)}{2\mu_0}d^3x$.
For a cluster in a static configuration (quite a good approximation for real systems) one
has
 \be
\frac{1}{2}\frac{d^2I}{dt^2}=0,
 \ee
from which eq.(\ref{eq.mvt}) derives.\\
In a more general derivation of the MVT (i.e., taking into account also surface integrals
in eqs.A7 and A8), the magnetic energy writes as
\be
U_B-\oint x_kF_{ij}dS_j= \frac{\phi^2}{r},
 \label{eq.UB}
 \ee
where $\phi \equiv \pi (B_*/\mu G) r_{vir}^2$ is the magnetic flux through the equatorial
section of the system. The trace of the surface integral in eq.(\ref{Pext}) writes as
 \be
 P_{ext}\oint(\vec{r}\cdot d\vec{S})=P_{ext}4\pi r_{vir}^3 \, ,
 \label{Pext.oint}
 \ee
and is usually considered in the analysis of the standard Virial
Theorem without the influence of a B-field (see, e.g., Carlberg et
al. 1997).\\
In the case of isothermal systems:
\be
2U=3\int_Vp_g~d^3x=3\int_Vc_s^2\rho_g d^3x\simeq 3c_s^2M_g
 \label{eq.tcost}
 \ee
where
 \be
c_s^2=\frac{k_BT_g}{\mu m_p}.
 \ee
The dark-matter particle kinetic energy writes as
\be
2K= \langle v_{dm}^2 \rangle \int_V\rho_{dm}d^3x = \langle v_{dm}^2 \rangle M_{dm} \, ,
 \label{eq.encin}
 \ee
where $ \langle v_{dm}^2 \rangle^{1/2}$ is the dark-matter particle velocity dispersion.
From eqs. (A.12)--(A.15) we obtain:
\begin{eqnarray}
 & 3\frac{k_BT_g}{\mu m_p}M_g + \langle v_{dm}^2 \rangle M_{dm} = & \nonumber \\
 & \xi \frac{GM_{vir}^2}{r_{vir}}\left(1-\frac{M_{\phi}^2}{M_{vir}^2 }+\frac{4\pi}{\xi
G}\frac{r_{vir}^4}{M_{vir}^2}P_{ext}\right) & .
 \end{eqnarray}
Since $1/\sqrt{3}  \langle v_{dm}^2 \rangle^{1/2} \approx c_s$, one obtains
\begin{eqnarray}
 & 3\frac{k_BT_g}{\mu m_p}(M_g + M_{dm}) = & \nonumber \\
 & \xi \frac{GM^2_{vir}}{r_{vir}}\left(1-\frac{M_{\phi}^2}{M_{vir}^2 }+\frac{4\pi}{\xi
G}\frac{r_{vir}^4}{M_{vir}^2}P_{ext}\right) & ,
 \end{eqnarray}
from which eq.(\ref{eq.TvirM}) follows setting $M_g + M_{dm}= M_{vir}$.

\end{document}